\documentclass[review,number,sort&compress]{elsarticle}

\usepackage{amssymb}
\usepackage{color}

\usepackage{lineno}
\usepackage{subfigure}

\begin{document}

\begin{frontmatter}


\title{Demonstration of polarization sensitivity of emulsion-based pair conversion telescope for cosmic gamma-ray polarimetry}


\author[kobe]{Keita Ozaki\corref{correspo1}}
\cortext[correspo1]{Corresponding author}
\ead{ozaki@radix.h.kobe-u.ac.jp}
\author[nagoya]{Satoru Takahashi\corref{correspo2}\fnref{note1}}
\cortext[correspo2]{Corresponding author}
\ead{satoru@radix.h.kobe-u.ac.jp}

\author[kobe]{Shigeki Aoki}
\author[kobe]{Keiki Kamada\fnref{note2}}
\author[kobe]{Taichi Kaneyama\fnref{note2}}
\author[kobe]{Ryo Nakagawa\fnref{note2}}
\author[kobe]{Hiroki Rokujo\fnref{note3}}

\address[kobe]{Kobe University, 3-11, Tsurukabuto, Nada-ku, Kobe, 657-8501, Japan}
\address[nagoya]{Nagoya University, Furo-cho, Chikusa-ku, Nagoya, 464-8602, Japan}

\fntext[note1]{Present address: Kobe University, 657-8501, Japan}
\fntext[note2]{Graduated.}
\fntext[note3]{Present address: Nagoya University, 464-8602, Japan}

\begin{abstract}
Linear polarization of high-energy gamma-rays (10 MeV-100 GeV) can be detected by measuring the azimuthal angle of electron-positron pairs and observing the modulation of the azimuthal distribution.
To demonstrate the gamma-ray polarization sensitivity of emulsion, we conducted a test using a polarized gamma-ray beam at SPring-8/LEPS.
Emulsion tracks were reconstructed using scanning data, and gamma-ray events were selected automatically.
Using an optical microscope, out of the 2381 gamma-ray conversions that were observed, 1372 remained after event selection, on the azimuthal angle distribution of which we measured the modulation.
From the distribution of the azimuthal angles of the selected events, a modulation factor of 0.21 + 0.11 - 0.09 was measured, from which the detection of a non-zero modulation was established with a significance of 3.06 $\sigma$.
This attractive polarimeter will be applied to the GRAINE project, a balloon-borne experiment that observes cosmic gamma-rays with an emulsion-based pair conversion telescope.
\end{abstract}


\end{frontmatter}


\section{Introduction}
\label{Introduction}
Cosmic gamma-ray observations can explore high-energy phenomena in space with imaging, timing, energy spectra, and polarization.
It is important to observe the polarization because the polarimetry of cosmic gamma-rays can be distinguished in the emission model of an astronomical object (e.g., \cite{Takata}).
However, high-energy gamma-ray polarimetry in space has not yet been investigated because of measurement difficulties \cite{Kotov}.

\begin{figure}[htbp]
\begin{center} \includegraphics[width=.5\textwidth]{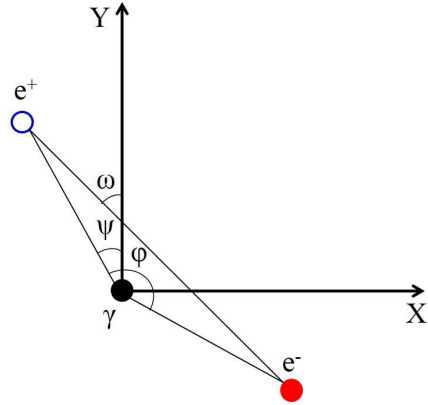} \end{center}
\caption{
Geometry of an electron-positron (e$^{-}$e$^{+}$) pair and the azimuthal angle in the detector plane.
The polarization direction is parallel to the Y-axis.
The angle $\phi$ is called the coplanarity angle. The angle $\psi$ is the azimuthal angle and $\omega$ is the observable azimuthal angle.
}
\label{geometry}
\end{figure}

Polarimetry with electron-positron pairs from high-energy gamma-rays was suggested in 1950 \cite{C.N. Yang}.
The signature of linearly polarized gamma-rays can be detected by measuring the azimuthal angle of the produced electron-positron pairs.
The azimuthal distribution can be presented as $\sigma$ = $\sigma$$_{0}$[1 + $P_{l}\cdot R\cdot\cos$(2$\psi$)], where $P_{l}$ is the linear polarization fraction, $R$ is the modulation factor, i.e., the maximum amplitude when $P_{l}$ is 100$\%$, and $\psi$ is the azimuthal angle between the polarization direction and the detected electron-positron pair.
Fig. \ref{geometry} shows the geometry of the produced electron-positron pair and its associated variables.
The angle $\omega$ between the polarization plane and the electron-positron plane is a directly observable parameter.

It is important to measure the azimuthal angle of an electron-positron pair using a thin converter before the tracks have traversed $\sim$10$^{-3}$ X$_{0}$ \cite{Kotov}, because the azimuthal angle is diffused by multiple Coulomb scattering while the particles traverse a detector.
However, it is also important to efficiently produce electron-positron pairs using a thick converter.
A pair polarimeter for linearly polarized gamma-rays has already been developed at the ground level \cite{C. de Jager}; however, this is not optimal for cosmic gamma-ray observations in terms of detection efficiency.
Several groups are studying the polarimetry for cosmic gamma-rays (e.g., \cite{D.Bernard}); however, it is difficult to develop a gas or solid-state detector capable of both suppressing multiple Coulomb scattering and converting gamma-rays efficiently.

Here we choose an other approach to develop an optimum pair polarimeter, keeping the high density of an emulsion, and using a sub-micron resolution.
\section{GRAINE project}
\label{GRAINE project}
The GRAINE project is a balloon-borne experiment that observes cosmic gamma-rays using an emulsion-based telescope \cite{GRAINEarxiv}.
The telescope comprises a nuclear photographic emulsion, i.e., a 3D-charged particle tracking detector with sub-micron spatial resolution.
Emulsion acts not only as a converter but also a tracker; therefore, it can detected the beginning of the electron-positron pair and measured its azimuthal angle before multiple Coulomb scattering could exert its deleterious effect.
The conversion efficiency of incoming gamma-rays can be enhanced by stacking emulsion films.
Therefore, the emulsion-based telescope can be optimum polarimeter for high-energy cosmic gamma-ray observations.

\begin{figure}[htbp]
\begin{center} \includegraphics[width=.82\textwidth]{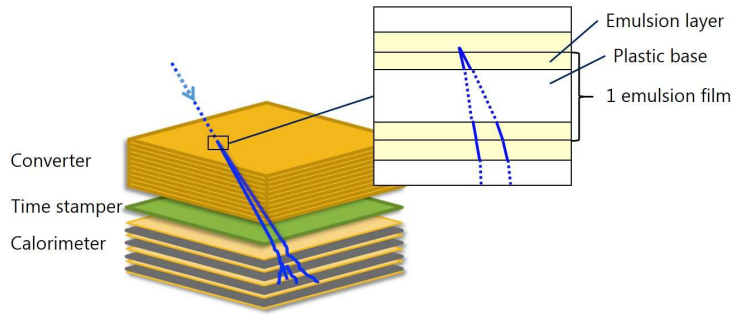} \end{center}
\caption{Schematic of the emulsion-based telescope.}
\label{EmulsionTelescope}
\end{figure}

Fig. \ref{EmulsionTelescope} shows the structure of the emulsion-based telescope.
The components (converter, time-stamper, and calorimeter) are aligned along the gamma-ray direction.

The first balloon-borne experiment (GRAINE-2011) was performed at the Taiki Aerospace Research Field in Japan in June 2011 \cite{Takahashi_GRAINE2011,Rokujo}.
The second balloon-borne experiment (GRAINE-2015) was performed at Alice Springs in Australia in May 2015 \cite{Takahashi_GRAINE2015,Ozaki}.
As a next step, GRAINE will convey a large-area emulsion-based telescope ($\sim$10 m$^{2}$) in repeated long-duration ($\sim$1 week) scientific balloon flights.

The telescope covers the energy range from 10 MeV to 100 GeV with high angular resolution (1.0$^{\circ}$ at 100 MeV, 0.08$^{\circ}$ at 1-2 GeV, obtained from both experimental data and simulation \cite{Takahashi_GRAINE2011}).
We aim to observe the polarization of astronomical objects such as pulsars, blazars, GRBs and gamma-ray flares.

In this paper, we report the first demonstration of polarization sensitivity of emulsion-based telescope.
\section{Experiment}
\label{Experiment}
\begin{figure}[htbp]
\begin{center} \includegraphics[width=.6\textwidth]{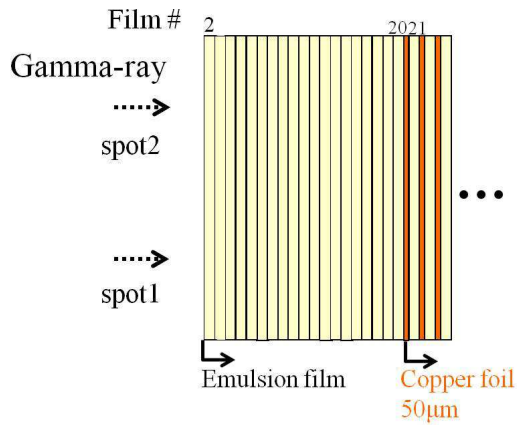} \end{center}
\caption{
Layout of the emulsion chamber. 
The OPERA films (12.5 cm $\times$ 10 cm $\times$ 293-$\mu$m thick) were stacked from film $\#$2 to $\#$20. 
A sandwich structure of the OPERA films and copper foils (50-$\mu$m thick) was stacked after $\#$20.
}
\label{emulsionchamber}
\end{figure}

Polarized gamma-ray beam exposure was performed at the SPring-8 facility in Japan using LEPS (Laser Electron Photon beam line at SPring-8) \cite{T.Nakano}.
The gamma-rays were produced by inverse Compton scattering of an Argon polarized laser (with a wavelength of 351 nm, 95$\%$ linear polarization) off the electrons (the energy was 8 GeV) circulating in the storage ring.
The maximum energy of the gamma-ray at the Compton edge was 2.4 GeV and the polarization fraction had a maximum value of 93$\%$ and 50$\%$ at 2.4 GeV and 1.5 GeV, respectively.
The produced gamma-rays were travelled to the experimental hutch located 69 m from the inverse Compton scattering region.
The gamma-ray beam was exposed perpendicular to the emulsion chamber for $\sim$1 s.
The beam spot size was collimated to a diameter of 2.3 cm.
The layout of the emulsion chamber is shown in Fig. \ref{emulsionchamber}.
In this experiment, we used OPERA films \cite{T.Nakamura}, which were jointly developed by the Nagoya University and the Fuji Photo Film Corporation.
Approximately 0.1$\%$ of the gamma-rays converted to the electron-positron pairs in one OPERA film.
\section{Analysis}
\label{Analysis}
\subsection{Gamma-ray event selection using scanning data}
After the emulsion films development, they were scanned by an automatic emulsion scanning system at the Nagoya University \cite{Morishima} in the 3$\times$3 cm$^{2}$ region around the beam center, and all the emulsion tracks were reconstructed \cite{K.Kodama}.
To enhance the polarization fraction of the gamma-ray beam, the connection window (the angle and position difference) of emulsion tracks between films was calculated in accordance with multiple Coulomb scattering, and the reconstruction threshold was set to 500 MeV/c.
We searched for unbiased single-tracks, which started at mid-film and connected to $\#$25, as candidates of gamma-ray events.
The search simultaneously required that the candidates be located in a 0.63 cm radius from the beam center.
This corresponded to the selection of $>$0.8 GeV for the primary gamma-rays.
Subsequently, 4228 gamma-ray event candidates were selected automatically from film $\#$5 to $\#$20.
\subsection{Manual checks and measurements using an optical microsope}
\label{Manual checks and measurements using an optical microsope}
The conversion points of each selected candidate were checked on a TV monitor; the microscopic view was projected via a CCD camera.
In this check, we confirmed that the selected candidates did not penetrate to the upper films, and 2551 events were confirmed to have a disappearance and then their azimuthal angles were measured when their distance was split by $>$0.5 $\mu$m.
Manually checked results of selected gamma-ray events is summarized in Table \ref{manualcheck_summary}.

The digitization of the pixel-pitch was 0.13 $\mu$m in this optical system.
The intrinsic resolution of the azimuthal angle based on the size of the silver crystal \cite{T.Nakamura} was calculated to be 2.8$^{\circ}$.
We also measured the arrival direction of the gamma-ray and the opening angle of the electron-positron pair.
\begin{table}[hbtp]
\caption{
Manually checked results of selected gamma-ray events. In total, the azimuthal angles of 2381 events were measured.
Penetrating tracks were contamination by through trakcs is due to the inefficiency of track recognition.
Some events did not split to the pair because one side track may be lower momentum and could not be found out.
Overlapped events were due to the same electron-positron pair event, but the different conversion point (1 up or down film) was selected between one side and the other side track.
}
\label{manualcheck_summary}
 \begin{center}
  \begin{tabular}{ccc}
	Selected events		&			&4228	\\\hline\hline
	penetrating		&			&1636	\\
	disappearance		&			&2551	\\
				&	split	&2381	\\
				&	not split	&170	\\
	overlapped		&			&41	\\
  \end{tabular}
 \end{center}
\end{table}
Gamma-ray beam had an energy dependence of polarization fraction naturally and included the secondary gamma-rays which were induced by residual gas of the beam pipe, so event selection was applied using the opening angle and the arrival direction of the electron-positron pair.
The opening angle correlates to the gamma-ray energy, so we applied the opening angle cut ($\theta_{\rm open}<$5 mrad was required) to purify the gamma-ray events.
To recover the high energy but uneven (large opening angle) events, we applied the cut of the arrival direction of the single track additionally, which selected on the large opening angle ($\theta_{\rm open}>$5 mrad) but the arrival direction is close ($<$1 mrad) to the beam center.
To estimate these selection efficiency, MC electron-positron pairs events from which gamma-rays converted in emulsion film were generated using Geant4 standard model (the recoil momentum were not taken into account), and the opening angle and the arrival direction distirbution of electron-positron pairs were obtained.
Consequently, the selection efficiency as a function of the gamma-ray energy was estimated to be 90$\%$ of the 2 GeV events but only 70$\%$ of the 1 GeV events by these selection.
\section{Results}
\label{Results}
\begin{figure}[htbp]
\begin{center} \includegraphics[width=.7\textwidth]{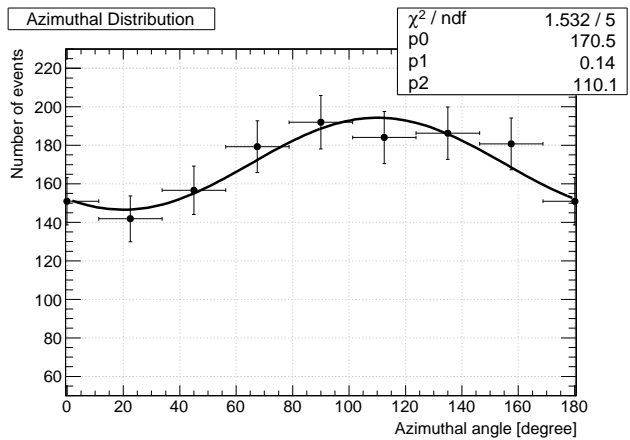} \end{center}
\caption{
The azimuthal distribution of the selected electron-positron pairs.
The fitted curve is N($\omega$) = p0$\cdot$(1+p1$\cdot$$\cos$(2($\omega$-p2))).
The free parameters p0, p1, and p2 indicate the baseline, amplitude, and polarization direction respectively.
The vertical error bar includes the statistical error only.
}
\label{azimuthdistribution}
\end{figure}

\begin{figure}[htbp]
\begin{center} \includegraphics[width=.7\textwidth]{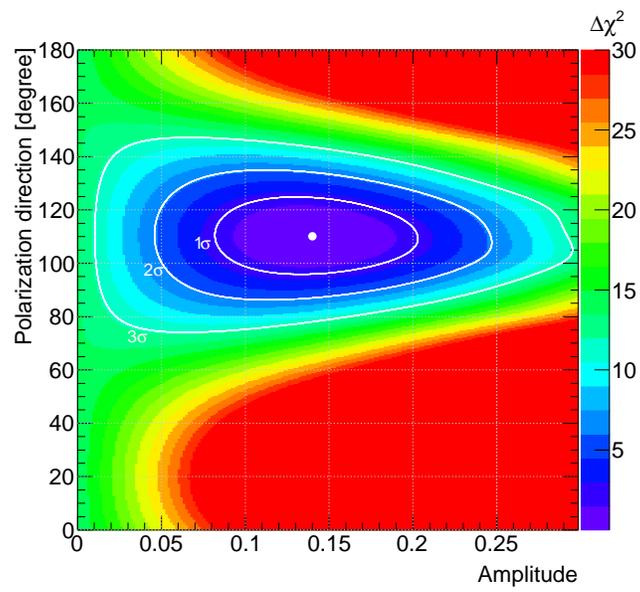} \end{center}
\caption{
$\Delta\chi^{2}$ map of the confidence contours in the amplitude-polarization direction plane.
The white dot represents the best-fit parameters, and we calculated the $\Delta\chi^{2}$ values relative to this point.
The null hypothesis (zero observed amplitude) can be ruled out with a 99.89$\%$ (3.06$\sigma$) confidence level.
}
\label{dchi_map}
\end{figure}

Fig. \ref{azimuthdistribution} shows the azimuthal distribution of the selected electron-positron pairs for the 1372 selected events.
The modulation curve can be seen in this result.
Fig. \ref{dchi_map} shows the $\Delta\chi^{2}$ map of the confidence contours in the amplitude-polarization direction plane.
The best-fit amplitude and polarization direction were 0.140 and 110.1$^{\circ}$, respectively.
The significance of the observed amplitude is at the 99.89$\%$ (3.06$\sigma$) confidence level.

To estimate the modulation factor $R$, we calculated the average polarization fraction $P_{\rm ave}$ as below.
\begin{equation}
{N^{\rm cut}(E) = N(E)\times \epsilon(E)}
\end{equation}
\begin{equation}
{P_{\rm ave} = \frac{P(E)\times N^{\rm cut}(E)}{\int N^{\rm cut}(E) dE}}
\end{equation}
$N(E)$ and $P(E)$ are the number of gamma-ray events and the polarization fraction as a function of the gamma-ray energy, according to \cite{T.Nakano}.
$\epsilon(E)$ is the selection efficiency discussed in Section \ref{Manual checks and measurements using an optical microsope}, 
and $N^{\rm cut}(E)$ is the number of selected gamma-ray events.
The measured $R\times P_{\rm ave}$ was estimated to be 0.14 + 0.07 - 0.06 (the error was calculated from $\Delta\chi^{2}$ = 2.3 contour in Fig. \ref{dchi_map}), and $P_{\rm ave}$ = 0.66, then the modulation factor $R$ was calculated to be 0.21 + 0.11 - 0.09.
\section{Conclusions}
\label{Conclusions}
We demonstrated the polarization sensitivity of an emulsion-based telescope, using a polarized gamma-ray beam at SPring-8/LEPS.
After event selection, the energy range is estimated to be 0.8-2.4 GeV.
From the distribution of the azimuthal angle of the 1372 selected events, taking into account the average polarization fraction of the beam of $P_{\rm ave}$ = 0.66 for the applied selection, but without making any correction for possible biases induced by that event selection, we obtain a modulation factor of 0.21 + 0.11 - 0.09.
The small opening angle selection affected to cut the energy uneven electorn-positron pair events and this cut may be enhanced the modulation factor, however, such uneven events are minor components in this experiment.
From this we exclude the hypothesis of a zero modulation factor, that is of a telescope insensitive to polarization, with a significance equivalent to more than 3 standard deviations.
%

\section*{Acknowledgements}
\label{Acknowledgements}
We appreciate the gamma-ray beam exposure support provided by the staff at SPring-8/LEPS.
We also appreciate the scanning support provided by the F-laboratory at the Nagoya University.
This work was supported by JSPS KAKENHI (Grant Numbers 20244031) and a Grant-in-Aid for JSPS Fellows.



\end{document}